\documentclass[article]{acmart}

\AtBeginDocument{%
  \providecommand\BibTeX{{%
    \normalfont B\kern-0.5em{\scshape i\kern-0.25em b}\kern-0.8em\TeX}}}

\copyrightyear{2023}
\acmYear{2023}
\setcopyright{acmlicensed}\acmConference[PRE]{Preprint}{May 2023}{NA}
\acmBooktitle{Preprint}
\acmPrice{}
\acmDOI{000}
\acmISBN{000}

\begin{document}

\title [The Barriers to Online Clothing Websites for Visually Impaired People]{The Barriers to Online Clothing Websites for Visually Impaired People: An Interview and Observation Approach to Understanding Needs}

\author{Amnah Alluqmani}
\affiliation{%
  \institution{The University of Sheffield}
  \streetaddress{2 Whitham Rd, Broomhall}
  \city{Sheffield}
  \country{UK}}
\email{asalluqmani1@sheffield.ac.uk }
\affiliation{%
  \institution{Umm Al-Qura
University}
  \city{Makkah}
  \country{Saudi Arabia}}
\email{asalluqmani@uqu.edu.sa }

\author{Morgan Harvey}
\affiliation{%
  \institution{The University of Sheffield}
  \streetaddress{2 Whitham Rd, Broomhall}
  \city{Sheffield}
  \country{UK}}
\email{ m.harvey@sheffield.ac.uk }

\author{Ziqi Zhang}
\affiliation{%
  \institution{The University of Sheffield}
  \streetaddress{2 Whitham Rd, Broomhall}
  \city{Sheffield}
  \country{UK}}
\email{ ziqi.zhang@sheffield.ac.uk }

\begin{abstract}
Visually impaired (VI) people often face challenges when performing everyday tasks and identify shopping for clothes as one of the most challenging. Many engage in online shopping, which eliminates some challenges of physical shopping. However, clothes shopping online suffers from many other limitations and barriers. More research is needed to address these challenges, and extant works often base their findings on interviews alone, providing only subjective, recall-biased information. We conducted two complementary studies using both observational and interview approaches to fill a gap in understanding about VI people's behaviour when selecting and purchasing clothes online. Our findings show that shopping websites suffer from inaccurate, misleading, and contradictory clothing descriptions; that VI people mainly rely on (unreliable) search tools and check product descriptions by reviewing customer comments. Our findings also indicate that VI people are hesitant to accept assistance from automated, but that trust in such systems could be improved if researchers can develop systems that better accommodate users' needs and preferences.

\end{abstract}



\maketitle

\section{Introduction}

Visually impaired or VI (i.e., blind and low vision) people cite shopping for clothes as one the most challenging everyday tasks they need to perform~\cite{lee2020challenges}. Advances in technology mean that VI people can now shop online without having to travel to stores, removing many of the physical challenges inherent in this task, and enhancing VI people’s lives ~\cite{j2021online}. As such, most VI people now engage in online shopping ~\cite{stangl2020person}.

However, shopping for clothes online introduces a number of new challenges and barriers for VI people, such as accessibility issues, an over-reliance on images, and scarcity of accurate clothing information~\cite{stangl2020person,wang2021revamp}. As such, many VI online shoppers have to rely on sighted assistance from friends and family. Unfortunately, most retailers rely heavily on product images to provide detailed information and, unlike sighted users, VI people need alternative text (alt text) to understand image content. However, in most cases, alt text is empty, lacks detail, or simply redirects to image paths~\cite{wang2021revamp}. Many online stores do not identify graphics (e.g., cartoon characters) on clothing products ~\cite{j2021online} or simply identify the characters without describing what they are doing, assuming that they are obvious from the image.

Existing studies addressing VI people's challenges in purchasing clothing online have relied primarily on interviews, which do not offer a holistic understanding of the problem. By combining observations with interviews, in this work we are able to obtain both objective information about specific issues encountered and participants' subjective, holistic online shopping experience. Observing participants conduct specific online shopping tasks can provide in-context evidence of the limitations of retailers’ websites and the opportunity to recall other challenges. Moreover, this approach produces rich data that can be used to generate design implications for creating new technology that meets the intended users' needs~\cite{rapp2021search}.

Furthermore, previous research has failed to explore VI people's challenges when attempting to coordinate outfits online and the services that help them to do so, which are not presently offered in online stores. No study has yet investigated how automated systems could help VI people select and coordinate outfits online and understand the services that they are more likely to trust automated systems to perform. To fill these gaps, both observational studies and interviews were conducted to investigate the clothing selection and coordination challenges of VI people shopping for clothes online. Their wishes regarding how automated systems could be designed to improve their experiences were also explored. A total of seventeen participants took part: ten in the observation tasks and fourteen in the interviews (seven participants joined both studies). We analysed the resulting data using thematic analysis.

Our findings corroborate many conclusions from prior literature but also contribute several new, previously unidentified challenges and potential solutions to help VI people shop for clothes online. This is the first comprehensive study of VI people's experiences in selecting and coordinating clothing online, with the opportunity to share their thoughts on how an automated system can be used to improve their online experience. It also makes a major contribution to research on designing better online assistance system for VI users. As well as reporting on our findings in detail, we translate these into design considerations, implications and recommendations to help make online shopping accessible and enjoyable for everyone.

\section{Literature Review}
In this section, we review prior research on three topics: offline clothes shopping challenges, online shopping challenges and online clothes shopping challenges.

\subsection{Offline Clothes Shopping Challenges}
Several studies have investigated the challenges VI people face when purchasing clothing offline. Lee et al.\cite{lee2020challenges} interviewed eight participants from the VI community about their clothes selection process, including the difficulties they face in confirming that they are selecting their desired product. The study revealed that the amount of product information impacts their ability to choose the correct item; a lack of product information makes it nearly impossible for VI people to find and compare items without assistance \cite{park2020understanding}.

Lack of information is not the only factor that impacts shopping self-sufficiency for VI people. Navigating inside shopping malls can be a stressful task~\cite{lee2020challenges}, especially when stores are poorly organised. Examples of prohibitive organisation include poorly laid-out department divisions, labels that lack colour consistency, very narrow aisles \cite{yu2015retail} and items in the wrong department. Further, stores often change their layouts, which can confuse VI people, who must often rely on their spatial memory from past visits~\cite{jones2019analysis}. When trying on clothes, VI people need to manage such challenges as closing the fitting room door; finding hooks for canes, purses or coats; looking for a mirror, and managing poor lighting. The payment process is also worth noting, as VI people often experience difficulty reading receipts, estimating their total before cashing out and using credit cards~\cite{lee2020challenges,yu2015retail}.

\subsection{Online Shopping Challenges}
VI people report that online shopping can give them a sense of reliability and control and is generally the preferred mode of shopping for clothes compared to the offline equivalent~\cite{ j2021online}. Nevertheless, online shopping comes with its own obstacles, including a lack of accessibility features in online spaces.

Table accessibility is an often-cited example of an online accessibility limitation. Web designers often use tables to present information in a compact manner; however, when screen readers are employed, a number of challenges arise, including the necessity for a good memory to remember each table cell's content and its relationship to the columns and rows. The more complex the table, the more difficult it is for screen reader users to associate cells with their columns and rows~\cite{whitney2020teaching}. Another difficulty is that screen readers read the text of tables without providing any sense of visual semantics; for example, screen readers do not provide any information regarding the use of common typographical signals, such as colour and font size~\cite{zimmerman2022recentering}.

Several guidelines for iOS applications~\footnote{https://developer.apple.com/accessibility/ios/.}, Android applications~\footnote{https://developer.android.com/guide/topics/ui/accessibility/} and web content~\footnote{http://www.w3.org/TR/WCAG10/} have been proposed to improve online accessibility for VI people. Unfortunately, most websites fail to adhere to such key VI accessibility criteria such as simple headings, high contrast ratios and consistent navigation. A study by Sohaib et al.~\cite{sohaib2017commerce} tested the accessibility of Australian business to consumer (B2C) e-commerce websites. Thirty B2C e-commerce websites were selected from the top sites in Australia. The selected websites were evaluated using A-Checker, an open-source web accessibility assessment system. The observed websites exhibited several accessibility problems that negatively impacted the experiences of VI visitors and were found to not adhere to even a basic level of web content accessibility criteria.

Seeking information also presents a challenge to VI online shoppers. One study compared the abilities of blind people and sighted individuals in accomplishing tasks online. The researchers~\cite{bigham2017effects} found that blind people take roughly twice as long as sighted people to complete an identical task. Even screen readers are often insufficient in helping a VI person make a purchasing decision. This is because of accessibility problems that are hard for the screen reader to handle. Such limitations include the generalisation of product descriptions~\cite{fuchs2012online} and insufficient details. The use of ambiguous language to describe products~\cite{kaufman2009understanding,wang2021revamp}, such as nonstandard colour names, can further inhibit VI people from recognising key product details. Further, online retailers rely on images to attract consumers instead of descriptive phrases ~\cite{yang2015design}. In some cases, product images rather than product descriptions contain the crucial information, or product images are not associated with alt-tags, which would allow screen readers to describe the content of the image to a VI shopper~\cite{sohaib2017commerce}.

\subsection{Online Clothes Shopping Challenges}
Shopping online for clothes specifically presents a set of additional challenges for VI users. For example, clothing websites often lack accessible information on size, colour, material, reviews, outfit coordination and age appropriateness ~\cite{j2021online,burton2012crowdsourcing,liu2019bought}.
Yang et al.~\cite{yang2015design} interviewed nine VI people about their challenges in purchasing clothes online. The participants noted that most retailers rely on images to present the features of their clothing and often provide no alternative or alt-tag descriptions. Williams~\cite{j2021online} reported that blind people are unhappy with how difficult it is to obtain clothing information online, such as colour, pattern, graphics and fit. The participants in both studies highlighted the challenge of understanding textual clothing descriptions. Some retailers employ complex words to name items or describe product details, such as using unfamiliar or culture-dependent colour names like ``teal'', ``sky blue'' and ``royal blue''~\cite{j2021online,wang2021revamp}.

Wang et al.~\cite{wang2021revamp} investigated the difficulties faced by VI shoppers when seeking clothing information online. They found a general lack of useful visual descriptions and a heavy reliance on images to provide information. To overcome these limitations, most VI participants sought human assistance when online shopping, while others paid for an online application to answer their questions. The participants reported reading customer reviews to better understand product details but still needed to rely on others to purchase clothing online.

Guanhong et al.~\cite{liu2019bought} conducted an in-depth study of how blind people shop online. They interviewed 20 blind people (face-to-face or by telephone), aged 17 to 48. The participants described wanting to be `normal' and to not stand out in mismatched clothing items. They wanted to be treated like sighted people but faced challenges in selecting and matching clothes online. According to Williams~\cite{j2021online}, VI shoppers look for information to help them decide whether a clothing item suits their style, personality and age. They also seek information to help them identify whether a pattern, colour or style will match another clothing item~\cite{stangl2018browsewithme}. Like any shopper, VI people seek information about the latest fashion trends~\cite{burton2012crowdsourcing} but face difficulties viewing other people's outfits to enrich their knowledge of fashion. VI people can complete most online shopping processes independently (e.g., making payments, returning products) but often require assistance for the key task of product selection.

Previous studies that have explored the challenges of clothes shopping for the VI people mainly relied on conducting interviews. However, observing VI people conducting shopping tasks provides objective and in-context evidence and demonstrates the challenges they face in a way that interviews cannot. Conducting an observational study also reminds the participants of challenges they might not have remembered during the interviews, thereby allowing us to mitigate recall bias and imperfect memory. Observing users also produces rich data that can be used to generate design implications for creating new technology that meets the intended users' needs~\cite{rapp2021search}.

A further point to note is that no previous studies have investigated the challenges VI people face in coordinating outfits online. To date, there is also a scarcity of research on how automated systems can be utilised to improve VI people's experiences. Interesting factors to consider include VI people's trust in automated systems, the kinds of automated services they would wish to have, and their fashion influences when purchasing or coordinating outfits. Research could also explore the offline environment to determine what services help VI people coordinate outfits in offline stores that are not offered so far in online stores.
In that context, to fill the aforementioned research gaps, we will address the following research questions:

\begin{description}
    \item[RQ1] What challenges do VI people face when independently selecting and coordinating outfits while online shopping?
    \item[RQ2] What type of assistance do VI people regularly make use of when shopping for clothes online?
    \item[RQ3] To what extent do VI people consider a specific celebrity or fashion style when selecting outfits?
    \item[RQ4]  What services are offered to assist VI people in coordinating outfits in offline stores that are not yet offered in online stores?
    \item[RQ5] What types of assistance would VI people like to have when coordinating outfits while shopping online?
\end{description}

\section{Method}
This work aims to understand VI people's behaviours and challenges while independently shopping for clothes online, including coordinating items. It also seeks to explore their influence on clothing preferences, the type of automated assistance they desire, and the level of trust they might place in such a system. 

To achieve these aims and address our research questions, we conducted two complementary studies. We employed task observations and standalone interviews to leverage the advantages both approaches and overcome some of the limitations inherent in employing each on its own. Task observation is effective and widely used in the HCI field for exploring potential design implications related to the development of a novel technology that meets the needs of its intended users ~\cite{rapp2021search}. While standalone semi-structured interviews are suitable for gaining a deep understanding of the intended users' views. Furthermore, the open-ended nature of the interview questions permits flexibility in the answers, allows unexpected responses to be elicited and recorded, and permits interviewees to discuss their perceptions of the task(s)~\cite{alshenqeeti2014interviewing}.

In Study 1, task observation was conducted to understand the participants’ in-context behaviours and identify barriers when performing specific shopping tasks. This approach allows the participants to recall and share examples of important details.  The tasks were framed in the context of a pre-defined simulated situation (similar to Borlund’s simulated work tasks\cite{borlund2000experimental}) to improve both internal and external validity by minimising confounding variables.  

Participants were asked to imagine that they are going to a party and need to buy a new outfit to wear to it, and to select one top and one bottom item of clothing, along with a third compatible item. During the task observations, participants completed multiple tasks, including sharing their screens, accessing the Amazon website (logged out of their accounts), and selecting and coordinating three clothing items. This retailer was chosen due to its popularity, worldwide ubiquity and inclusion of various product-related information, such as product descriptions, images and reviews. Moreover, the implementation of the pre-planned tasks was restricted to a single retailer to minimise confounding or extraneous variables, thus simplifying the analysis. After completing the tasks, we conducted short follow-up interviews to ask questions about their experiences purchasing clothes online and the tasks they completed during the observation.

While Study 1 focused on obtaining an understanding of VI people’s behaviours when purchasing clothes online, Study 2 gathered narrative data about their perspectives and needs. In Study 2, the participants were asked questions via standalone semi-structured interviews about their outfit selections, coordination processes, requirements and needs.

The task observations and the follow-up interviews in Study 1 lasted around 50 minutes in total - approximately 30 minutes for observation and 20 minutes for the follow-up interview - and the interviews in Study 2 lasted around 30 minutes. Due to Covid-19, both studies were conducted remotely using video conferencing applications, except for one participant, who preferred to participate via a phone call. We audio-recorded the interviews in both studies, and video recorded (by capturing only the screen) the observation tasks to ensure that all relevant data was gathered \cite{jamshed2014qualitative}. Throughout the task observations, we made notes in relation to our research questions and later used the recordings to make further notes.

\subsection{Participant Recruitment}
A total of 17 participants took part in this research (see Table \ref{tab:demographicInfo} for demographic information), including seven who participated in both studies. Overall, 10 participants took part in both parts of Study 1 (i.e., task observation and follow-up interview), and 14 were interviewed for Study 2. Participant recruitment was performed by our contacts at several UK charities, who advertised the studies on their social media accounts and in their online groups. Interested people contacted us via email, and we confirmed whether they met the criteria (explained later in this section). If the potential participant confirmed that he/she met the criteria, then we accommodated his/her preference to take part in one or both studies. Those who were willing to participate in both studies were able to choose which study to complete first based on their availability and preferences. We continued the recruitment process until we reached data saturation.

We applied the same three recruitment criteria for both studies: aged over 18, the absence of other disabilities aside from visual impairment and blindness (to exclude other challenges that may have affected purchasing behaviour, such as hearing or physical impairment), and prior experience with shopping online (at least once). To avoid any accessibility issues when providing consent, the participants had the option of verbally providing their consent before the interview (an audio recording was made to maintain integrity) or filling in the consent forms and returning them by email. All participants received a £10 Amazon voucher for participating in each study, resulting in remuneration at or above the UK living wage~\footnote{See http://livingwage.org.uk for more information}.

\subsection{ Research Procedures}
\begin{figure}[H]
    \centering
    \includegraphics[scale=0.4]{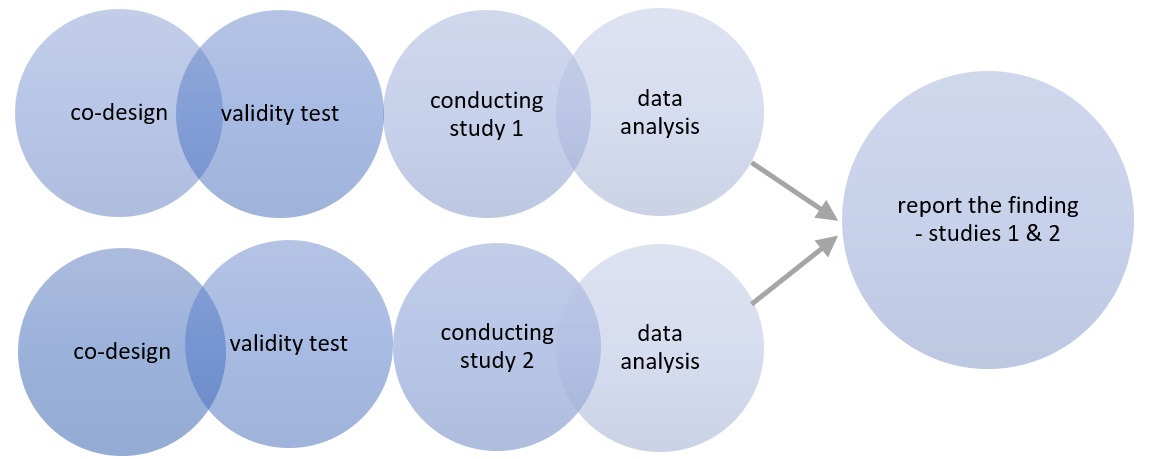}
    \Description[Overview flow diagram of the research procedures]{The research was conducted as follows, with some overlap:
    Co-designing, validation, conducting the main study, and analysing the data.
    These steps were implemented independently, but in the last step, the findings of both studies were merged as they were complementary.}
    \caption{Overview flow diagram of the research procedures: co-design with the Sheffield Royal Society for the Blind (SRSB), validity test, pilot studies, main studies, data analysis, and reporting the results. The research steps were implemented independently, but in the last step, the findings of both studies were merged as they were complementary.}
    \label{fig:flow_diagram}
\end{figure}

As Figure~\ref{fig:flow_diagram} shows, we conducted two complementary studies. Both studies were co-designed in collaboration with representatives from the Sheffield Royal Society for the Blind (SRSB), a local society for VI people. We started our collaboration with the SRSB by discussing VI people's challenges and their experiences of communicating and working with VI people. Then, we discussed the initial proposals for study designs and explored whether they were suitable. For instance, were the tasks achievable by individuals with visual impairments, and would they require any support to complete the tasks? Alongside communicating with the charity, the members of our team also validated the study designs in relation to our aims and research questions. This process was implemented iteratively.

The interview design was inspired by the interview protocol refinement framework~\cite{castillo2016preparing}, whereby the interview is implemented in four steps. In the first step, the validity of the interview questions is tested by aligning them with the research questions. The researchers then, in the second step, set protocols for conducting the interviews as an inquiry-based conversation. As part of this, the questions must adhere to spoken language conventions and everyday practices so they are simple to understand and respond to. It is also essential to follow the rules of social conversation when designing these protocols. Examples of such social rules are asking one question at a time and using introductory questions, transitions between questions, and closing questions. Preparing a draft interview script can assist with developing such protocols and thereby smoothing the interview conversation. An interview script provides written information that directs the interviewee during the interview; it includes information the participant should know and guides them on where the conversation should be heading. After building the interview questions and protocols in these first two steps, in the third step, the interview protocols are then reviewed, with the team members providing feedback and suggestions on how to enhance reliability. In the final step, the interview is piloted on a small sample of participants to check the functionality of the developed interview protocols, and appropriate amendments are made to resolve any issues identified.

The testing criteria in this pilot study were inspired by the criteria highlighted by Dikko \& Maryam~\cite{dikko2016establishing} and Hennink et al.~\cite{hennink2020qualitative}. These criteria are described as follows:
\begin{itemize}
     \item Identifying any difficulty in implementing the proposed tasks in the task observation method 
     \item	Identifying any ambiguity in the interview questions 
      \item Determining the expected time for completing each of the observations, subsequent interviews, and the separate standalone interviews.
      \item	Checking whether the proposed tasks and questions cover all aspects of the research questions 
       \item Testing the task observation and the interview guides.
\end{itemize}

After applying this framework and thereby confirming the quality of both study designs, we then conducted the studies without a strict order (e.g., we did not wait to complete study 1 before beginning study 2) to accommodate the preferences of the participants. Each participant was randomly assigned an ID to protect their identity. Analysis of the data from completed studies was done iteratively, with the new data from each additional completed study being Incorporated into the analysis, allowing us to assess when saturation had been reached. The coding and analysis of data from the two studies were performed separately, and then the findings of both studies were reported together as they were complementary.

\subsection{Study Involving Participants with Visual Impairments}
It is important to point out that conducting studies with people who have visual impairments presents a number of additional challenges and requires the researcher to be more attentive to participants' well-being during the studies. We faced additional obstacles as we conducted our interviews and observational tasks using online video conferencing applications. Some applications suffered from poor accessibility, and some VI people expressed discomfort because they were unfamiliar with them. Thus, providing familiar applications or alternative platforms, such as phone calls for the interviews, was essential.

Conducting the observational tasks online with VI people was challenging. They not only had to deal with the accessibility limitations of being online but also had to perform other tasks, such as sharing the screen. This required them to interact with several dialogs, which were mostly inaccessible to their screen readers. Some potential participants wanted to take part but were reluctant to perform the online shopping tasks, not only because of accessibility issues but also because they did not want to perform them in front of others to avoid being judged for their behaviour.

It was also vitally important to ensure the pre-planned tasks were realistic for and achievable by the population. We did this by co-designing the study with people who have direct contact with the research population (i.e., representatives from the SRSB).

\begin{table*}[ht]
  \caption{Participant demographic information}
   \Description[Participant demographic information]{Participant demographic information; 17 participants took part across both studies. 
   The table shows the ID, age group, perception level, visual experience, type of assistance system, and completed study number of each participant.
       
    Participant 1F is female in the 18–24 age group, blind with congenital visual experience, uses screen readers, and participated in Study 2.
   
   Participant 1C is Male in the 25–34 age group, blind with acquired visual experience, uses Voice assistance(Siri) and screen readers, and participated in Studies 1 and 2.
   
    Participant 1D is female in the 35-44 age group, blind with congenital visual experience, uses screen readers, and participated in Studies 1 and 2.
   
   Participant 1G is female in the 55-64 age group, Low Vision with congenital visual experience, uses Magnification tool	, and participated in Studies 1 and 2.
   
  Participant 1L is female in the 45-54 age group, blind with congenital visual experience, uses Screen readers, and participated in Study 2.
  
    Participant 1N is Male in the 25-34 age group, Blind with acquired visual experience, uses Screen readers, and participated in Study 2.
    
   Participant 1Q is female in the 45-54 age group, Low Vision with Congenital visual experience, uses colour contrast settings	, and participated in Studies 1 and 2. 
       
    Participant 1Q is female in the 75 or more age group, Low Vision with Congenital visual experience, uses Large cursor setting, and participated in Studies 1 and 2.

    Participant 1S is female in the 35-44 age group, Blind with Congenital visual experience, uses Magnification tool, and participated in Studies 1 and 2.
    
    Participant 1F is male in the45-54 age group, Blind with acquired visual experience, uses Voice assistance, and participated in Study 2.    
    
    Participant 2D is Female in the 35-44 age group, Blind with acquired visual experience, uses Screen readers (Voiceover), and participated in Study 2.    
    
  Participant 2E is male in the 45-5 age group, Blind with Congenital visual experience, uses Screen readers, and participated in Study 2.
  
  Participant 2Q is Female in the 45-54 age group, Blind with Congenital visual experience, uses Screen readers, and participated in Study 2.     

  Participant 2H is male in the 45-54 age group, Blind with Congenital visual experience, uses Screen readers(Voiceover), and participated in Study 2.  
  
  Participant 1E is female in the 35-44 age group, Blind with Congenital visual experience, uses Screen readers(Voiceover), and participated in Study 1.  
  
  Participant 1A is male in the 55-64 age group, Blind with Congenital visual experience, uses magnification tool, and participated in Study 1.   
 
  Participant 1B is female in the 45-54& age group, Blind with Congenital visual experience, uses magnification tool, and participated in Study 1.   
    
   }
  \label{tab:demographicInfo}
  \begin{tabular}{cccccccl}
    \toprule
  ID&Age Group&Gender&Perception level&Visual Experience&Type of assistance system&Stud[y/ies]\\
    \midrule
     1F&18-24&Female
     &Blind
     &Congenital
     &Screen readers
     &2 \\

  1C&25-34&Male	
  &Blind
  &Acquired	
  &Voice assistance(Siri)
  \& screen readers	
  &1 \& 2 \\
  
  1D&35-44&Female	
  &Blind	
  &Congenital	
  &Screen readers	
  &1 \& 2 \\
  
  1G&55-64&Female&Low Vision	&Congenital	
  &Magnification tool
  &1 \& 2 \\
  
  1L&45-54&Female	
  &Blind	
  &Congenital	
  &Screen readers	
  &2\\
  
  1N&25-34&Male&Blind
  &Acquired	
  &Screen readers
  \&Voice assistance
  &2\\
  
  1Q&45-54&Female&Low Vision	&Congenital	
  &colour contrast settings	
  &1 \& 2\\

  1M&75 or more&Female&Low Vision	&Congenital	 
  &Large cursor setting
  &1 \& 2\\
  
  1S&35-44&Female&Blind
  &Congenital
  &Magnification tool
  &1 \& 2\\
  
  2F&45-54&Male
  &Blind
  &Acquired	
  &Voice assistance	
  &2\\
  
  2D&35-44&Female	
  &Blind 
  &Acquired	
  &Screen readers (Voiceover)	
  &1 \& 2\\
  
  2E&45-54&Male
  &Blind 
  &Congenital	
  &Screen readers	
  &2\\
  
  2Q&45-54&Female
  &Blind
  &Congenital	
  &Screen readers	
  &2\\
  
  2H&45-54&Male
  &Blind 
  &Congenital	
  &Screen readers (Voiceover)	
  &2\\
  
  1E&35-44&Female
  &Blind&Congenital
  &Screen readers (Voice-over)
  &1\\
  
  1A&55-64&Male	
  &Blind
  &Congenital	
  &magnification tool	
  &1\\
  
  1B&45-54&Female
  &Blind
  &Congenital
  &magnification tool	
  &1\\

    \bottomrule
  \end{tabular}
\end{table*}

\section{Data Analysis}
Our analysis process was inspired by those used by Braun and Clarke~\cite{braun2006using}. We began data familiarisation by transcribing the interviews using a partially-automated technique, repeatedly checking the transcripts and field notes against the audio and video recordings. We read and revised the collected data for clarity, and we annotated and revised the field notes while watching the video recordings of the participants’ shopping tasks.

The early steps in the task observations followed Eriksson et al.’s~\cite{eriksson2015qualitative} suggested analysis method, including asking ourselves what the field notes indicate and what interesting and unique things are related to our research questions. However, while conducting subsequent observational studies and reading more into the data, we began to notice interesting behavioural patterns, including the process followed when searching for desired items and the sources of the descriptions relied upon (e.g., product titles, descriptions, customer reviews). We also identified the behaviours that occurred when participants encountered challenges that hindered their ability to make purchasing choices, and we noted the issues experienced with the tools they relied upon when conducting shopping activities, such as the limitations of the Amazon search engine from the perspective of VI users.

Then, with the qualitative data analysis software NVivo, we used an open coding techniques to develop and modify the codes as we proceeded through the coding process. To ensure reliability, the team members discussed the codes and themes several times, iteratively revising the codes and checking them with each new transcript to see if new ones should be generated. We also reviewed the codes to ensure they were related to the research questions~\cite{braun2012thematic}.

The standalone interviews were analysed separately from the task observations, but the deeper the analysis went, the more we found strong connections between the codes generated from each. We therefore clustered the codes in each study and ended up with the same five themes. We report the final themes and the findings of both studies as a complementary story in the following section.

\begin{figure}[ht]
    \centering
    \includegraphics[width=\linewidth]{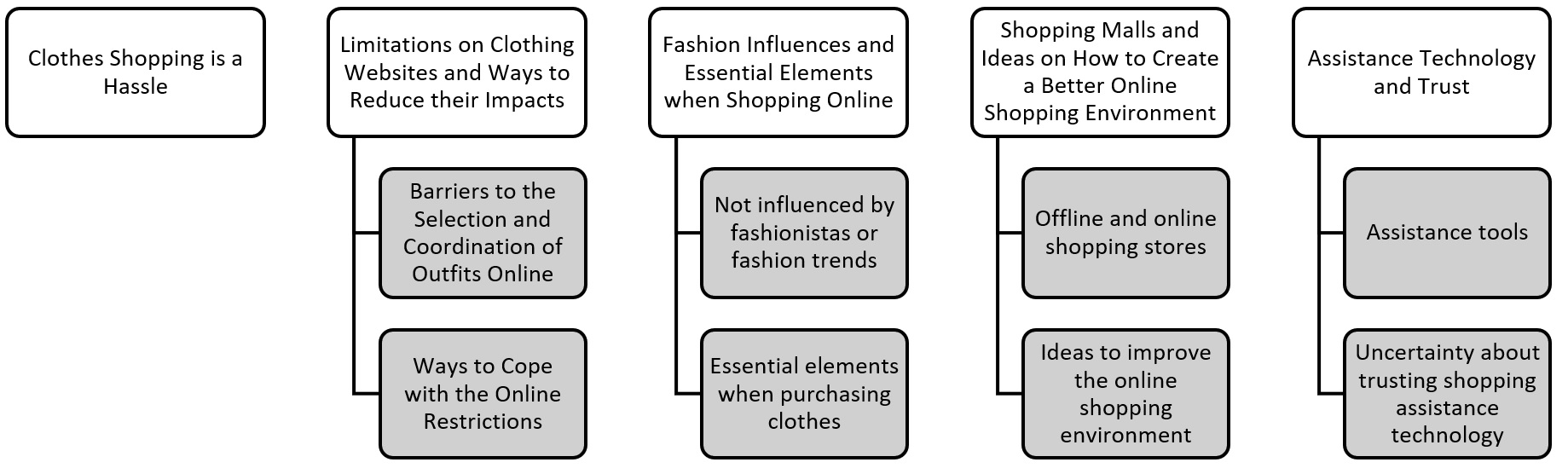}
    \Description[Diagram of themes]{Five key themes emerged from our analysis of studies 1 and 2.
    
    The first theme is Clothes Shopping is a Hassle.
    
    The second theme is Limitations on Clothing Websites and Ways to Reduce their Impacts. There are two codes generated from the second theme. 
    The first code is Barriers to the Selection and Coordination of Outfits Online. 
    The second code is Ways to Cope with the Online Restrictions.
    
    The third theme is Fashion Influences and Essential Elements when Shopping Online. There are two codes generated from the third theme.
    The first code is Not influenced by fashionistas or fashion trends. 
    The second code is Essential elements when purchasing clothes.
    
    The fourth theme is Shopping Malls and Ideas on How to Create a Better Online Shopping Environment. There are two codes generated from the fourth theme.
    The first code is Offline and online shopping stores.
    The second code is Ideas to improve the online shopping environment.
    
    The fifth theme is Assistance Technology and Trust. There are two codes generated from the fifth theme. The first code is Assistance tools. The second code is Uncertainty about trusting shopping assistance technology.}
    \caption{Diagram of themes identified from Studies 1 and 2}
    \label{fig:codesGraph}
\end{figure}

\section{Findings}
Five key themes emerged from our analysis of both studies (see figure \ref{fig:codesGraph}); a total of 59 individual items resulted from Study 2, and 34 items from Study 1.
Although the semi-structured interviews in Study 2 generated in-depth knowledge, the observational studies uncovered several aspects that did not emerge from the interviews. For instance, it revealed the inconsistency between how the participants represented their process of purchasing clothing online and their actual online shopping behaviour. Examples of observed behaviour include the prominent use of the search tool in the purchasing process and the trust in customer reviews over retailer descriptions when making a purchase choice. In addition, conducting purchasing tasks in Study 1 helped the participants bring evidence of the challenges they mentioned in the interviews and recall new ones, such as the restrictions of the search tool and the usage of conflicting terms when describing clothing items.

In this section, we discuss each theme, cite illustrative participant quotes, and refer to incidents noted during the task observations.

\subsection{Clothes Shopping is a Hassle}
When discussing the difficulty of clothes shopping, participants compared it to grocery and electronics purchasing. Participants 1N and 1M noted that buying clothes requires multiple visual tasks, such as identifying colours, styles, sizes, and materials, and that fashion trends change, making shopping in this context much harder.

For example, participant 1C said that he felt he could trust any sighted person for help buying food``if I just want some food, I could call any friend if I didn't have time to go online. However, with clothes, I want someone I can trust [1C, Study2]''. Although interviewees believed that online shopping eliminates the many difficulties associated with physical stores, they described it as a ``guessing game'' that was overwhelming when it involved shopping for clothes.

\subsection{Limitations on Clothing Websites and Ways to Reduce their Impacts}

\subsubsection{Barriers to the Selection and Coordination of Outfits Online}
In response to RQ1 ({\em What challenges do VI people face when independently selecting or coordinating outfits while online
shopping?}), participants outlined many challenges that render their ability to select and coordinate clothing items online more difficult, including the restrictions of clothing descriptions and the unreliability of the current descriptions and tools. \\

Several limitations of clothing descriptions were brought up in relation to online shopping.
Participants identified the lack of style descriptions as one of the most frequent constraints. They sought details of clothing items, including details about sleeves, pockets, buttons and zippers. In the case of an item having buttons, for instance, VI people wanted to know,``How are the buttons? Are they press stud? Are they crest buttons? [1G, Study2]''. Such information is almost never provided in the descriptions of items online.

The lack of adequate graphic descriptions was also identified in Study 2 and noted during the observational tasks. Participants indicated that retailers rarely provide descriptions of graphics and designs printed on clothes that would enable customers with VI to mentally imagine how it looks. Participant 1C referenced a T-shirt item he chose during the observation tasks, which was described as `Take this with you on your summer holiday'. Commenting on this, he stated that this description was insufficient for a person with VI to predict the graphic. Instead, he suggested a better description, such as ``a picture of the camper van in a green field with whatever cloudless sky [1C, Study2]''. To obtain such a description, a VI shopper would need to rely on a sighted friend or relative. 

How a piece of clothing with a  graphic looks is about more than just what is represented by the graphic. For example, as reported by participant 1C, how the graphic on a clothing item feels (e.g., soft print) and the position of the graphic is also important. Participant 1S, who is congenitally blind, provided an example of how difficult it is to figure out the position of a graphic:``It might say a heart-printed t-shirt, but sometimes the heart can be quite abstract because you have to hunt to find out where it is on the top and especially if it's a similar colour as the rest of the t-shirt [1S, Study2]''. Participant 1L also provided the case of an item with a tie-dye design. She simply wanted to know whether the design started on the right or left side of the item.

In addition to the lack of style and graphic information, colour descriptions were also highlighted by the participants as being either missing or described using ambiguous terms. For example, the use of terms like `deep blue', `sapphire', `champagne' or `charcoal', or uniformly described, such as grey, blue or pink, which do not provide a clear indication of the colour shade. VI people are interested in not only the colours or colour shades of an item, but also the position and extent of each color. Talking about this issue, 1L said: ``Quite often, it will just say blue and white, but it wouldn't tell you [if] it's blue and white stripes or blue with white flowers or, you know, blue and white spots[...] I don't know which one is most... which one has the most colour [2Q, Study2]''.

Unclear sizing information was another issue identified with clothing descriptions. This includes information such as the item's overall length, the length of its sleeves, or how loose or tight it is. Interviewees wanted to know the size of the other elements on the clothing items, such as pockets and buttons. One participant, 1L, described how challenging it is to identify size when retailers use the term `one size', as this does not provide any information about which size group the item is for.

Another reported problem was the unreliability of the meta-information about the clothing and, subsequently, of the Amazon search tool, which uses this data to filter items. When discussing existing online descriptions, the participants used terms such as \emph{deceptive} and \emph{misleading}. Participant 1L shared her experience of encountering inaccuracy in the descriptions of pockets, which are her essential clothing item elements. She prefers pockets with fastenings because they give her more security when she takes her phone, money and keys outside. She reported her experience of ordering yoga fleece bottoms:``It said zip fastenings for the pockets. When I got it, it didn't have a zip fastening! It just had standard pockets that were on the side! [1L, Study2]''. 1L also shared details about a time she bought a dress online. Although she received a dress with the same basic design as the one described on the website, it was in the wrong colours: ``It was a black dress with red flowers on it, and the description was a white dress with black flowers [1L, Study2]''.

\begin{figure}
  \centering
    \includegraphics[width=.7\linewidth,height=170pt]{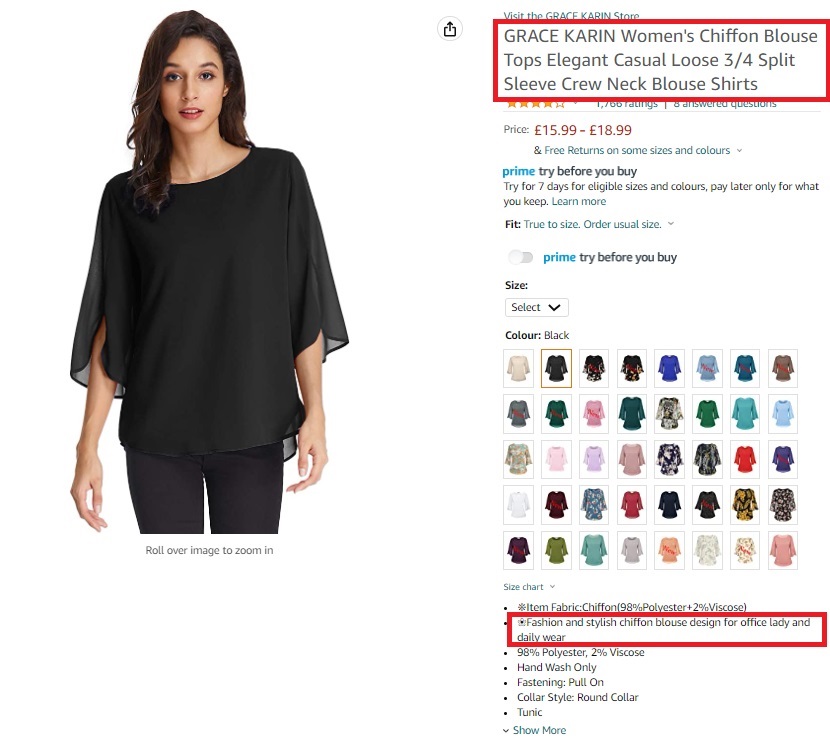}   \Description[An example of using contradictory terms]{An example of using contradictory terms (i.e., casual and office wear) when describing clothing items online. The title of the item is GRACE KARIN Women's Chiffon Blouse Tops Elegant Casual Loose 3/4 Split Sleeve Crew Neck Blouse Shirts. The description section describes the item as a Fashion and stylish chiffon blouse design for office lady and daily wear.}
  \caption{An example of using contradictory terms (i.e., casual and office wear) when describing clothing items online.}\label{fig:unreliableInfo}
\end{figure}

In another related issue, participants also noted misclassification of items. For example, sometimes clothing was labelled as casual when it should be formal, and vice versa. The observational task revealed such a case of contradictory information. Participant 1D selected a blouse described as `Chiffon Blouse Tops Elegant Casual', thinking it was more like party wear. However, when she checked the description, it said `office wear' (see figure  \ref{fig:unreliableInfo}), so she decided to try a different item. She commented on this experience, stating that ``if I went into every single [item's descriptions], I’d be here all day and all night [1D, Study1]''. This is another clear example of an issue that is considerably less problematic for sighted individuals, who can either identify the misclassification error from the images, or can quickly skim the description. This is considerably more frustrating and time-consuming for shoppers who have to rely on a screen reader.

The Amazon search tool was also reported as being frequently unreliable. We observed that VI people rely mainly on the search feature on Amazon to look for their desired items. Although this search capability saves time and effort when filtering multiple items, the participants found it challenging to review the resulting items because they were very likely to receive extensive results, including entirely unrelated items. For example, participant 1M typed ``ladies' smart shoes'' to look for a third item compatible with her previous selections. She noted that the search engine suggested items related to the word `shoes' and not the entire search query. This is another instance where the time cost of such imprecision for sighted users would be minimal but for those who have to rely on screen readers it is considerable.

\subsubsection{Ways to Cope with the Online Restrictions}
Participants highlighted several approaches they take to cope with the current limitations. One such technique is developing familiarity with particular websites to reduce the possibility of making mistakes. One of the interviewees commented on this technique: ``If I go to a new shop I’m not familiar with, then it can become sort of really finding my way around and not knowing what to get [1S, Study2]''. The obvious downsides to this is a reluctance to branch out from familiar sites and brands, and a need to refamiliarise oneself if the website changes its design.

Another technique that appeared in both the interviews and observational tasks was that the participants preferred to stick to dark colours, mainly black, which they called ``\emph{safe colours}''. The term arose from the subjects' belief that dark colours match all other colours. In addition, to minimise the problems of matching outfits, most participants preferred to select plain items (i.e., with no patterns or graphics). One of the respondents commented:``I tend to always wear the same colour: top and bottom. So it’s safe, I suppose. I would say I’m a quite safe shopper [2D, Study2]''.

The observational tasks also revealed that most of the subjects confirmed their choices by reading the reviews. Although this might also be true for sighted people, we noted that the VI people often did this to fill in the missing information about the selected product and validate the `official' descriptions. When the participants were asked about the type of information they searched for when examining review data, they mainly mentioned sizing, length measurements and quality, and seeking other customers’ general opinions. They trusted customer comments more than the descriptions provided by retailers. This was observed when participant 2D relied on customer descriptions, despite the fact that their descriptions directly contradicted those of the retailer. She selected a top that was described as a casual blouse and a cardigan described as a beach cardigan. However, in the reviews, she found that the items were considered more formal. Another participant, 1S, who uses magnification tools when browsing the internet, finds reading review data overwhelming; she described her method of scanning the review data:``I tend to [...] look at the latest ones [reviews] first and try and get a feel. Yeah. Because from that, you can generally normally get a feel for the item. And then I would move this across to see the different star ratings. And then quite often I'd look at the very best and the very worst might I get a good idea of what the issues are [1S, Study1]''.

\subsection{Fashion Influences and Essential Elements when Shopping Online}
In response to RQ3 ({\em To what extent do VI people consider a specific celebrity or fashion style when selecting outfits?}), most of our participants stated that fashionistas, influencers or fashion styles do not typically influence their clothing choices. Instead, participants had other concerns, including how comfortable the item is or whether it includes certain practical elements, such as pockets. Some participants mentioned following specific brands, colour trends, or refer to their body shapes or to other people around them and what they wear. 

The results also revealed that the lack of visual ability and fashion resources most likely affect the participants' fashion influences. This was exemplified by participant 1N, who has an acquired sight problem. He expressed how following fashion styles became less attractive when he became blind: ``When I was sighted, I was wondering quite a lot, but not now, […] Now it’s kind of a regular thing [1N, Study2]''. Another participant ID, argued that fashion sources such as magazines and TV shows do not really accommodate VI users. She commented: ``A lot of the fashion information is about photos, and there's just not enough description for me to want to take an interest [1D, Study2]''.

On the other hand, some participants have their own influences when it comes to fashion and clothing selection. Participant 1M likes to follow trends in colours rather than styles:``I probably follow the colours that [...] may be prevalent this year [1M, Study2]''. In another example, participant 1Q, who is partially sighted, loves fashion and follows fashionistas. However, she prefers to shop based on her body shape, rather than following specific fashion trends. 

Other participants stated that the main factors when selecting clothing items are comfort and ease of access. For example, participant 1F, a young blind participant, stated that for her outfits are functional rather than aesthetic. When asked about following fashionistas or fashion styles, she laughed and said: ``I don't tend to look for things to try to look better or to try to impress other people with what I look like or try to copy a [...] celebrity style [1F, Study2]''.

\subsection{Shopping Malls and Ideas on How to Create a Better Online Shopping Environment}
With regard to RQ4 ({\em What services are offered to assist VI people in coordinating outfits in offline stores that are not yet offered in online stores?}) and RQ5 ({\em What types of assistance would VI people like to have when coordinating outfits while shopping online?}), the participants stated how offline shopping services impacted their ability to coordinate clothing items when compared to online shopping. They also discussed how automated assistance systems could improve the experience for VI shoppers.

Our findings indicate that the tactile and physical properties of offline stores, including the opportunity to feel and try on items, provide the ability for VI people to better coordinate clothing items when compared to purchasing items online. In online shopping, however, it was noted that coordinating items is done mainly ``by coordinating in multiple windows [1S, Study2]'', which can be hard on the eyes for people with limited vision as it requires focusing in and out multiple times. Moreover, this process is time-consuming. To avoid this, it was observed that participants mainly chose simple styles with dark colours or picked items based on colour coordination.

The participants also discussed how automated systems may be used to improve their experience in online shopping; below, we discuss these findings.

\paragraph{Detailed descriptive information}
The participants asked about the necessary requirements to improve online shopping and mentioned the need for better quality descriptive information, suggesting that clothing items should be described in greater detail. Participant 1D expressed her expectation for the availability of descriptions online: ``if you pick a dress [...], it will say, What colour is that? [...] what's the pattern like? [...] Is it polka dots? or is it just squares, zigzags? [1D, Study2]''.

While offering detailed textual information could be beneficial for avid users of screen readers, it could be tiring for others. To combat this, participant 2F suggested offering an audio button for verbal descriptions. Participant 2D felt that having clothing measurements in textual form, rather than in tables would help them to shop. 

Another issue raised by the participants was enhanced colour descriptions. They suggested standardising colour palettes and associating them with codes or putting them on a consistent numerical scale. Participant 1M described how colour information can be presented: ``If there are 45 blues for example and it would have a code that would tell you that it's blue which is this 'X241' and that's your colour code. [...] the Pantone colours [1M, Study2]''.

\paragraph{Clothing and outfit recommendation systems}
Another important service requested by the participants was recommender systems. The majority of the participants sought a recommendation system that offered item coordination, rather than just suggesting similar ones. Participant 2F said:``If there was a way of being able to almost put two images together and the computer's telling you whether they matched or not, [...] Are they the same colour or are they matching colours or matching styles. If there was a way that the computer could tell you about that [it] would be fantastic [2F, Study2]''. 

Another participant, 1Q, transitioned from partially sighted to blind. When she coordinates outfits online, she tries to balance the style, such as a loose top with a tight bottom or vice versa. She also coordinates items based on colour agreement. However, she finds this process tiring, so she suggested a system that might take three of her selected items and put them together so she could see whether they are well coordinated.

Others suggested a system that alerts users to poor matches, rather than helping them to choose coordinating items. Such a system should alert customers when selected items do not match or are not appropriate for a specific occasion. Participant 2F expressed how these recommendations might be helpful: ``It could tell [the automated system] that light green and bright orange don’t really go together [... and that] having stripes one way and stripes the other way [is] not a good idea [2F, Study2]''. This recommendation approach was generally preferred by the participants over recommending items that go together: ``if it was telling you what to put together, it’s a bit more restrictive [2D, Study2]''. 

In addition to these suggested recommendation systems, a system that narrows down search results based on previous purchases was also suggested: ``Perhaps if it keeps track of what you've already purchased and then picks a similar style clothes that cuts out things, [...] you might not be interested in. Um, because, you know, there are several things I never wear.[...], if that would be removed from the search engine [1S, Study2]''. An important consideration here would be whether to consider short-term items (e.g., those currently in the user's basket) or longer-term items, like those that the user had bought in recent weeks or months.

\paragraph{What should recommender systems know?}
The participants shared particular suggestions on how recommendation systems could be made more effective for their needs. They stated that it would be useful if the system recommended outfits based on customers' colour preferences and body shapes. Participant 1M, for example, suggested that the system should recommend items based on the user's colour season, such as whether they are an autumn, summer, winter or spring colour. For instance, the system could clarify that the colour white is not suitable for autumn colour. Another participant, 1Q, suggested a system that recommends items based on skin colour, as some colours are more complimentary toward certain skin tones than others. Participant 2D, on the other hand, suggested a system that recommends items based on the user's body shape, such as whether they have a pear or apple-shaped body. This is another good example of something that a sighted person might be able to determine from product images but which would be essentially hidden from an IV user.

\subsection{Assistance Technology and Trust}
With regard to RQ2 ({\em What type of assistance do VI people regularly make use of when shopping for clothes online?}), participants reported that they use screen readers, automated voice assistants (e.g., Alexa and Siri), and colour contrast and magnification tools to access the Internet. When asked about their trust in automated assistance, they stated that they are hesitant to trust such systems, instead placing more trust in sighted people (e.g., friends or relatives) when purchasing clothing. They believe that trust needs to be established based on experience. They are happy to use new assistance systems, but they do not trust them when selecting expensive clothing or clothing meant for special occasions. Participant 1C, who is an early adopter of new technology and a technophile, commented: ``I would certainly buy an average day-to-day outfit, summer clothes holiday clothes. [...] I would spend average prices [...], £23 maximum each for an item [...] using the automated system to begin with [1C, Study2]''. This implies an expectation that using the automated system would lead to poor choices the user might regret. 

The participants explained their reasons for not trusting the automatic systems to help in their clothes purchasing process; the main issue raised was that most automatic systems do not know their customers' specific needs or style preferences. Participant 1C, for example, discussed the importance of building a system that knows its customers and provides services based on customers' preferences. 1C shared her thoughts about automatic systems:``They [automated systems] don't know your needs. They don't know your preferred type of clothes. So they're not really going to help me in terms of telling me [...] if this meets [...] my specific needs. If this is going to fit me [1F, Study2]''.

Another related point that explains the participants' mistrust is that automated systems do not have the ability to determine the suitability of clothing items for different occasions, for example a beach party or formal dinner.  Automated systems also do not have the ability to predict how a piece of clothing might fit a user. Participant 1F commented: ``a visual person, especially like a family member that I live with every day knows how big I am. [...] they will know what size is likely to fit me [1F, Study2]''.

Participants were also more confident relying on human assistance because sighted humans can describe garments by comparing them to other items already known to or purchased by the VI shopper. Participant 1D commented: ``what helps for me is they're able to say: `Oh, it looks like that top that you've got', so they can reference it something that I actually have at home. So it makes it more real [1D, Study2]''. This might be possible in a recommender system by considering a user's purchase history and would be a form of explainable recommendation.

\section{Discussion}
Our findings demonstrate the challenges VI people face when shopping on clothing websites, what influences their clothing preferences, what types of automated assistance they seek and what level of trust they might place in such a system. In this section, we discuss our findings and their contributions to the existing literature.

The initial objective of this work was to identify the barriers to using online clothing websites for VI people. Our results indicate that the scarcity of clothing descriptions is a key challenge, which is consistent with the findings in the literature \cite{wang2021revamp,liu2019bought,j2021online, stangl2018browsewithme,yang2015design}. Our findings reveal that applying the observational approach provides a holistic understanding of the issues under investigation. This approach allowed the participants to provide prompt examples of the issues they encountered when shopping for clothes online, leading to better discussions. Our findings also indicate that the observational tasks disclose challenges that were not reported during the interviews. For example, our observational study reveals several behavioural patterns, including validating the retailer's description with review data and using the search function as a primary tool for findings items online, thereby suggesting ideas on how to offer better solutions. It supports the idea that using task observation in concert with interviews produces rich data and insights that may not be identified from one of these methods alone. Such insights can be used to inform design implications for creating new technology~\cite{rapp2021search}.

Our observational studies uncovered a previously unreported issue — deceptive or incorrect product descriptions and the knock-on effect on the reliability of the Amazon search tool — because in most cases, it led to unrelated items. However, there is a risk of bias, as this study was conducted exclusively using the Amazon website. One interpretation of the limitations outlined in this observational study is that the Amazon website consists of many retailers from various countries, which might explain the use of unfamiliar terms. For example, the word `blouse' is used to describe items that are referred to as `shirts' in some countries. However, retailers may also attempt to purposely mislabel items to gain sales that they would otherwise miss.

To cope with the limitations of online clothing websites, the results show that VI people achieve confidence in their selection by confirming their choices with sighted, trusted individuals or by reading customer reviews. These findings are in line with previous research \cite{stangl2018browsewithme,wang2021revamp,liu2019bought}. However, in contrast with previous studies \cite{j2021online,wang2021revamp}, the results from this research did not report the use of chat services as assistance tools (e.g., Aira or Seeing AI). The reason for the lack of uptake of these support options among the study population is unclear, but may relate to the fact that most of the participants were blind (only three participants had low vision). This research, however, revealed that, potentially as a result of missing out on such support options, the participants restricted their shopping strategies to minimise risk (e.g., by selecting only black or plain-coloured items or by only visiting familiar websites). One issue that emerged from these findings was the participants' impressions of restriction and inflexibility when selecting clothing items online.

Our findings indicate that our participants were not interested in following specific designs, fashion models or influencers; instead, they follow specific brands, colour trends or refer to their individual body shapes. This might help us understand the most essential aspects of designing a personalised system for VI people. As the sample population of this study was of various ages with differing degrees of interest in fashion, it is possible that this result is valid for other populations.

Our results confirm that VI people are generally hesitant to trust automated systems, which is in line with the literature \cite{wang2021revamp}. One significant finding explaining their trust in humans is that humans can reference items already owned or known to the VI shopper, which would be helpful in visualising an image of how the new item looks.

Despite this scepticism, the investigation of possible systems to assist in coordinating and selecting clothing online revealed that VI people would be interested in improved assistance systems, including image captioning and recommendation systems. Our results have significant implications for designing such an assistance system for VI people (details of the implications for design will be discussed in the following subsection).

The contradiction between the necessity for automated systems and the reluctance to trust them leads to an essential discussion of the types of assistance systems that VI people are most likely to trust. According to these results, together with the findings of previous work \cite{wang2021revamp,stangl2018browsewithme}, it seems that VI people would welcome new assistive technology. However, they put a high degree of trust in systems that emphasise the probability of error \cite{macleod2017understanding} or systems that are more personalised \cite{lee2022imageexplorer}. It is unclear whether VI people would trust a system to recommend clothing items based on objective features, such as colour, size, and pattern \cite{gharaei2021content} or more aesthetic qualities, such as judging style \cite{yu2018aesthetic}. 

\subsection{Limitations}
In order to minimise confounding variables, our participant sample excluded VI people with other disabilities or comorbidities. A second limitation was that, for practical data collection reasons, the observational tasks were conducted exclusively using the Amazon website. Although most participants were already familiar with Amazon and considered it to be a more accessible website than many others, and the use of a single site reduced experimental complexity, the task should be repeated on a range of different websites with varying levels of perceived accessibility and participant familiarity. Another limitation is that observing people doing something can also impact their actions. Despite these limitations, the present work has enhanced our knowledge of possible design implications for developing a clothing shopping assistance system.

\subsection{Design Implications}
Our findings have a number of important design implications for future practice.

\subsubsection{Developing Systems to Automatically Produce More VI People–Friendly Captions for Clothing Images}

The main constraint VI people face in enjoying independent online shopping is the absence of clothing descriptions. Providing adequate clothing descriptions will enhance knowledge of the available products, thereby strengthening the ability to build a more reliable assistance system.
Although the lack of clothing descriptions has been reported in previous literature~\cite{wang2021revamp,liu2019bought,j2021online, stangl2018browsewithme,yang2015design} and many studies have attempted to offer solutions, such as developing an automatic fashion captioning system ~\cite{tateno2020method}, our findings indicate that the scarcity of clothing descriptions remains valid.

Our findings also suggest that briefly describing products does not assist VI individuals in imagining how they look. It is therefore important to develop systems that can produce more VI-friendly captions tailored to the needs of VI people instead of extant approaches, which are tailored to sighted users. One possible design that could help develop more suitable captions for VI people is to specify crucial clothing characteristics for such users. This includes mentioning the presence of graphics, buttons, zippers and pockets on clothing items as well as providing descriptive information about these features, such as their size and kind (e.g., an invisible zipper, press-stud buttons or a patch pocket). Methods have been proposed that can control the length of generated captions~\cite{deng2020length}, forcing the model to produce more detailed descriptions. Other work has emphasised the importance of providing users with captions that realistically present the level of uncertainty/inaccuracy in the automated descriptions~\cite{macleod2017understanding}.

\subsubsection{Validate Clothing Product Descriptions}
Our results show that contradictory or deceptive descriptions on online clothing websites make it challenging for the participants to select their desired items. The contradictions appeared in the descriptions themselves as well as between how the retailers described the product and how customers described the same product in their reviews. One solution may be to filter product information by validating the information in the product descriptions with the review data. This can be done by providing additional information regarding the accuracy level of the product information and giving the VI shoppers the chance to make a more informed decision.

\subsubsection{Colour Identification and Scaling System}
Our participants reported restricting their choices when purchasing clothing on their own to predominantly dark colours (mainly black) to reduce the chance that they will receive undesirable or incompatible colours. This is not necessarily the colour they would want to choose but is simply a safe choice. Future research might develop a colour identification and scaling system (e.g., assigning 01 as a metric number to the darkest blue colour, 02 to the lighter blue colour and 10 to the lightest blue colour) to provide a more flexible shopping environment for VI people and broaden their shopping choices. Such a system could analyse product images to identify colours and then clearly describe the item's colours using a recognised colour scheme/palette (e.g., blue, red, and green), each associated with specific scaling numbers. Another potential solution might be for the system to outline the percentage of each colour in a piece of clothing.
These suggested systems would permit VI people to imagine how dark or light the selected item would be, helping them to match items better. It would also allow them to determine how similar or different the selected item is from the items they already own and the extent of each colour in a clothing item.

\subsubsection{Cut Out Unlikely-to-Purchase Items from Search Engine Results}
As shown in our observational study, the wide range of item results from the Amazon search engine introduces further challenges for VI people. One design idea is to develop a system that tracks users' shopping behaviours to understand their preferences (similar to existing recommender systems). The user preference information can then be used to curate the search engine results by excluding items unlikely to be purchased and presenting only user-preferred items. The system could also personalise the results based on the user's fashion influences. This design suggestion may also be of benefit to sighted users.

\subsubsection{Fashion Warning System}
Existing limitations of clothes shopping websites make it challenging for VI people to select their desired items successfully. Our results show that VI people mainly stick to familiar websites and restrict their selections to plain, dark clothing to avoid purchasing incompatible or undesirable items. As an implication of these results, it would be a good idea to develop a system that works as a monitor, warning its users when they make undesirable selections. For example, the system might warn users if they match casual items with formal ones or attempt to coordinate tops with busy patterns with bottoms with a different and clashing pattern. Such a system could be controlled by understanding the users themselves better. For example, the system could obtain information about the users’ skin colour tone, their favoured colour season (e.g., spring or autumn colour ranges) or their body shape (e.g., apple or pear shape). With this information, the system would be able to generate personalised warnings.

\subsection{Future Work}
To gather more widely applicable findings, future research should consider a broader sample of the population and attempt to understand whether and how various other disabilities interact with visual impairment for such tasks. Additional observational tasks should be repeated on a range of different websites with varying levels of perceived accessibility and prior participant familiarity. As VI people often get assistance from a sighted person when shopping for clothes~\cite{wang2021revamp}, further research could look into how the online purchasing behaviour of VI people differs in the presence of sighted assistance. For future researchers, an observational approach will provide a complete view of online issues. However, researchers should consider VI people's possible difficulties when completing the tasks and offer alternative tools if VI participants express discomfort.
Another interesting topic that might be considered in future research is exploring the factors that affect the level of reliability of assistance systems (e.g., question-answering and recommendation systems) for VI people.

\section{Conclusions and Summary of Contributions}
In this study, a complementary approach (interviews and observations) was used to obtain insights into the selection and coordination challenges of VI people shopping for clothes online. Their ideas for design changes to automated systems in order to improve their experiences were also examined, contributing to research on the design of VI-user-friendly online assistance systems.

The study design enabled us to directly observe behaviours and the challenges faced in context, rather than relying on potentially unreliable or biased recall, providing useful context that prompted the participants to recall and share examples not raised during interviews alone. For instance, when using Amazon's search tool during the observational task, they encountered problems such as misleading titles and the results including unrelated items, which made it harder to efficiently find the desired item. Further challenges included contradictory descriptions or customer comments that were inconsistent with the descriptions (e.g., using both `casual' and `formal' to describe clothing items). Observing the participants as they conducted online shopping tasks demonstrated the difficulties that VI users face when searching for and assessing desired items; avoiding deceptive and unclear information; and coordinating outfits. Observing these challenges helped us identify VI users' actual needs, which is essential for developing useful design recommendations, and allowed the participants to suggest useful ideas for how an automated system might help.

The standalone interviews revealed several interesting fashion-related factors that may influence VI people's clothing selections, such as brands, body shapes and colour trends, instead of fashionistas or style trends. Another important contribution of this research is the improved understanding of the types of automated systems that VI people seek in clothing shopping websites and are likely to trust.

The combination of interviews and observational studies generated novel findings that led to several design suggestions, including systems to automatically produce clothing image captions that are more friendly to VI people, validated clothing product descriptions, colour scaling systems and fashion warning systems. Our results show the necessity for further research to understand the challenges VI people face and the need for a better online shopping experience. Researchers can benefit from the challenges discussed in the present study to run future studies and develop more accessible and usable technological solutions.

\newpage
\bibliographystyle{ACM-Reference-Format}
\bibliography{Main.bib}


\begin{thebibliography}{33}


\ifx \showCODEN    \undefined \def \showCODEN     #1{\unskip}     \fi
\ifx \showDOI      \undefined \def \showDOI       #1{#1}\fi
\ifx \showISBNx    \undefined \def \showISBNx     #1{\unskip}     \fi
\ifx \showISBNxiii \undefined \def \showISBNxiii  #1{\unskip}     \fi
\ifx \showISSN     \undefined \def \showISSN      #1{\unskip}     \fi
\ifx \showLCCN     \undefined \def \showLCCN      #1{\unskip}     \fi
\ifx \shownote     \undefined \def \shownote      #1{#1}          \fi
\ifx \showarticletitle \undefined \def \showarticletitle #1{#1}   \fi
\ifx \showURL      \undefined \def \showURL       {\relax}        \fi
\providecommand\bibfield[2]{#2}
\providecommand\bibinfo[2]{#2}
\providecommand\natexlab[1]{#1}
\providecommand\showeprint[2][]{arXiv:#2}

\bibitem[Alshenqeeti(2014)]%
        {alshenqeeti2014interviewing}
\bibfield{author}{\bibinfo{person}{Hamza Alshenqeeti}.}
  \bibinfo{year}{2014}\natexlab{}.
\newblock \showarticletitle{Interviewing as a data collection method: A
  critical review}.
\newblock \bibinfo{journal}{\emph{English linguistics research}}
  \bibinfo{volume}{3}, \bibinfo{number}{1} (\bibinfo{year}{2014}),
  \bibinfo{pages}{39--45}.
\newblock


\bibitem[Bigham et~al\mbox{.}(2017)]%
        {bigham2017effects}
\bibfield{author}{\bibinfo{person}{Jeffrey~P Bigham}, \bibinfo{person}{Irene
  Lin}, {and} \bibinfo{person}{Saiph Savage}.} \bibinfo{year}{2017}\natexlab{}.
\newblock \showarticletitle{The Effects of" Not Knowing What You Don't Know" on
  Web Accessibility for Blind Web Users}. In
  \bibinfo{booktitle}{\emph{Proceedings of the 19th international ACM SIGACCESS
  conference on computers and accessibility}}. \bibinfo{pages}{101--109}.
\newblock


\bibitem[Borlund(2000)]%
        {borlund2000experimental}
\bibfield{author}{\bibinfo{person}{Pia Borlund}.}
  \bibinfo{year}{2000}\natexlab{}.
\newblock \showarticletitle{Experimental components for the evaluation of
  interactive information retrieval systems}.
\newblock \bibinfo{journal}{\emph{Journal of documentation}}
  \bibinfo{volume}{56}, \bibinfo{number}{1} (\bibinfo{year}{2000}),
  \bibinfo{pages}{71--90}.
\newblock


\bibitem[Braun and Clarke(2006)]%
        {braun2006using}
\bibfield{author}{\bibinfo{person}{Virginia Braun} {and}
  \bibinfo{person}{Victoria Clarke}.} \bibinfo{year}{2006}\natexlab{}.
\newblock \showarticletitle{Using thematic analysis in psychology}.
\newblock \bibinfo{journal}{\emph{Qualitative research in psychology}}
  \bibinfo{volume}{3}, \bibinfo{number}{2} (\bibinfo{year}{2006}),
  \bibinfo{pages}{77--101}.
\newblock


\bibitem[Braun and Clarke(2012)]%
        {braun2012thematic}
\bibfield{author}{\bibinfo{person}{Virginia Braun} {and}
  \bibinfo{person}{Victoria Clarke}.} \bibinfo{year}{2012}\natexlab{}.
\newblock \bibinfo{booktitle}{\emph{Thematic analysis.}}
\newblock \bibinfo{publisher}{American Psychological Association}.
\newblock


\bibitem[Burton et~al\mbox{.}(2012)]%
        {burton2012crowdsourcing}
\bibfield{author}{\bibinfo{person}{Michele~A Burton}, \bibinfo{person}{Erin
  Brady}, \bibinfo{person}{Robin Brewer}, \bibinfo{person}{Callie Neylan},
  \bibinfo{person}{Jeffrey~P Bigham}, {and} \bibinfo{person}{Amy Hurst}.}
  \bibinfo{year}{2012}\natexlab{}.
\newblock \showarticletitle{Crowdsourcing subjective fashion advice using
  VizWiz: challenges and opportunities}. In
  \bibinfo{booktitle}{\emph{Proceedings of the 14th international ACM SIGACCESS
  conference on Computers and accessibility}}. \bibinfo{pages}{135--142}.
\newblock


\bibitem[Castillo-Montoya(2016)]%
        {castillo2016preparing}
\bibfield{author}{\bibinfo{person}{Milagros Castillo-Montoya}.}
  \bibinfo{year}{2016}\natexlab{}.
\newblock \showarticletitle{Preparing for Interview Research: The Interview
  Protocol Refinement Framework.}
\newblock \bibinfo{journal}{\emph{Qualitative Report}} \bibinfo{volume}{21},
  \bibinfo{number}{5} (\bibinfo{year}{2016}).
\newblock


\bibitem[Deng et~al\mbox{.}(2020)]%
        {deng2020length}
\bibfield{author}{\bibinfo{person}{Chaorui Deng}, \bibinfo{person}{Ning Ding},
  \bibinfo{person}{Mingkui Tan}, {and} \bibinfo{person}{Qi Wu}.}
  \bibinfo{year}{2020}\natexlab{}.
\newblock \showarticletitle{Length-controllable image captioning}. In
  \bibinfo{booktitle}{\emph{Computer Vision--ECCV 2020: 16th European
  Conference, Glasgow, UK, August 23--28, 2020, Proceedings, Part XIII 16}}.
  Springer, \bibinfo{pages}{712--729}.
\newblock


\bibitem[Dikko(2016)]%
        {dikko2016establishing}
\bibfield{author}{\bibinfo{person}{Maryam Dikko}.}
  \bibinfo{year}{2016}\natexlab{}.
\newblock \showarticletitle{Establishing Construct Validity and Reliability:
  Pilot Testing of a Qualitative Interview for Research in Takaful (Islamic
  Insurance).}
\newblock \bibinfo{journal}{\emph{Qualitative Report}} \bibinfo{volume}{21},
  \bibinfo{number}{3} (\bibinfo{year}{2016}).
\newblock


\bibitem[Eriksson and Kovalainen(2015)]%
        {eriksson2015qualitative}
\bibfield{author}{\bibinfo{person}{P{\"a}ivi Eriksson} {and}
  \bibinfo{person}{Anne Kovalainen}.} \bibinfo{year}{2015}\natexlab{}.
\newblock \bibinfo{booktitle}{\emph{Qualitative methods in business research: A
  practical guide to social research}}.
\newblock \bibinfo{publisher}{Sage}.
\newblock


\bibitem[Fuchs and Strauss(2012)]%
        {fuchs2012online}
\bibfield{author}{\bibinfo{person}{Elisabeth Fuchs} {and}
  \bibinfo{person}{Christine Strauss}.} \bibinfo{year}{2012}\natexlab{}.
\newblock \showarticletitle{Online shopping involving consumers with visual
  impairments--a qualitative study}. In \bibinfo{booktitle}{\emph{International
  Conference on Computers for Handicapped Persons}}. Springer,
  \bibinfo{pages}{378--385}.
\newblock


\bibitem[Gharaei et~al\mbox{.}(2021)]%
        {gharaei2021content}
\bibfield{author}{\bibinfo{person}{Narges~Yarahmadi Gharaei},
  \bibinfo{person}{Chitra Dadkhah}, {and} \bibinfo{person}{Lorence Daryoush}.}
  \bibinfo{year}{2021}\natexlab{}.
\newblock \showarticletitle{Content-based clothing recommender system using
  deep neural network}. In \bibinfo{booktitle}{\emph{2021 26th International
  Computer Conference, Computer Society of Iran (CSICC)}}. IEEE,
  \bibinfo{pages}{1--6}.
\newblock


\bibitem[Hennink et~al\mbox{.}(2020)]%
        {hennink2020qualitative}
\bibfield{author}{\bibinfo{person}{Monique Hennink}, \bibinfo{person}{Inge
  Hutter}, {and} \bibinfo{person}{Ajay Bailey}.}
  \bibinfo{year}{2020}\natexlab{}.
\newblock \bibinfo{booktitle}{\emph{Qualitative research methods}}.
\newblock \bibinfo{publisher}{Sage}.
\newblock


\bibitem[Jamshed(2014)]%
        {jamshed2014qualitative}
\bibfield{author}{\bibinfo{person}{Shazia Jamshed}.}
  \bibinfo{year}{2014}\natexlab{}.
\newblock \showarticletitle{Qualitative research method-interviewing and
  observation}.
\newblock \bibinfo{journal}{\emph{Journal of basic and clinical pharmacy}}
  \bibinfo{volume}{5}, \bibinfo{number}{4} (\bibinfo{year}{2014}),
  \bibinfo{pages}{87}.
\newblock


\bibitem[Jones et~al\mbox{.}(2019)]%
        {jones2019analysis}
\bibfield{author}{\bibinfo{person}{Nabila Jones},
  \bibinfo{person}{Hannah~Elizabeth Bartlett}, {and} \bibinfo{person}{Richard
  Cooke}.} \bibinfo{year}{2019}\natexlab{}.
\newblock \showarticletitle{An analysis of the impact of visual impairment on
  activities of daily living and vision-related quality of life in a visually
  impaired adult population}.
\newblock \bibinfo{journal}{\emph{British Journal of Visual Impairment}}
  \bibinfo{volume}{37}, \bibinfo{number}{1} (\bibinfo{year}{2019}),
  \bibinfo{pages}{50--63}.
\newblock


\bibitem[Kaufman-Scarborough and Childers(2009)]%
        {kaufman2009understanding}
\bibfield{author}{\bibinfo{person}{Carol Kaufman-Scarborough} {and}
  \bibinfo{person}{Terry~L Childers}.} \bibinfo{year}{2009}\natexlab{}.
\newblock \showarticletitle{Understanding markets as online public places:
  Insights from consumers with visual impairments}.
\newblock \bibinfo{journal}{\emph{Journal of Public Policy \& Marketing}}
  \bibinfo{volume}{28}, \bibinfo{number}{1} (\bibinfo{year}{2009}),
  \bibinfo{pages}{16--28}.
\newblock


\bibitem[Lee et~al\mbox{.}(2022)]%
        {lee2022imageexplorer}
\bibfield{author}{\bibinfo{person}{Jaewook Lee}, \bibinfo{person}{Jaylin
  Herskovitz}, \bibinfo{person}{Yi-Hao Peng}, {and} \bibinfo{person}{Anhong
  Guo}.} \bibinfo{year}{2022}\natexlab{}.
\newblock \showarticletitle{ImageExplorer: Multi-Layered Touch Exploration to
  Encourage Skepticism Towards Imperfect AI-Generated Image Captions}. In
  \bibinfo{booktitle}{\emph{Proceedings of the 2022 CHI Conference on Human
  Factors in Computing Systems}}. \bibinfo{pages}{1--15}.
\newblock


\bibitem[Lee et~al\mbox{.}(2020)]%
        {lee2020challenges}
\bibfield{author}{\bibinfo{person}{Jihyun Lee}, \bibinfo{person}{Jinsol Kim},
  {and} \bibinfo{person}{Hyunggu Jung}.} \bibinfo{year}{2020}\natexlab{}.
\newblock \showarticletitle{Challenges and Design Opportunities for Easy,
  Economical, and Accessible Offline Shoppers with Visual Impairments}. In
  \bibinfo{booktitle}{\emph{Proceedings of the 2020 Symposium on Emerging
  Research from Asia and on Asian Contexts and Cultures}}.
  \bibinfo{pages}{69--72}.
\newblock


\bibitem[Liu et~al\mbox{.}(2019)]%
        {liu2019bought}
\bibfield{author}{\bibinfo{person}{Guanhong Liu}, \bibinfo{person}{Xianghua
  Ding}, \bibinfo{person}{Chun Yu}, \bibinfo{person}{Lan Gao},
  \bibinfo{person}{Xingyu Chi}, {and} \bibinfo{person}{Yuanchun Shi}.}
  \bibinfo{year}{2019}\natexlab{}.
\newblock \showarticletitle{" I Bought This for Me to Look More Ordinary" A
  Study of Blind People Doing Online Shopping}. In
  \bibinfo{booktitle}{\emph{Proceedings of the 2019 CHI Conference on Human
  Factors in Computing Systems}}. \bibinfo{pages}{1--11}.
\newblock


\bibitem[MacLeod et~al\mbox{.}(2017)]%
        {macleod2017understanding}
\bibfield{author}{\bibinfo{person}{Haley MacLeod}, \bibinfo{person}{Cynthia~L
  Bennett}, \bibinfo{person}{Meredith~Ringel Morris}, {and}
  \bibinfo{person}{Edward Cutrell}.} \bibinfo{year}{2017}\natexlab{}.
\newblock \showarticletitle{Understanding blind people's experiences with
  computer-generated captions of social media images}. In
  \bibinfo{booktitle}{\emph{proceedings of the 2017 CHI conference on human
  factors in computing systems}}. \bibinfo{pages}{5988--5999}.
\newblock


\bibitem[Park et~al\mbox{.}(2020)]%
        {park2020understanding}
\bibfield{author}{\bibinfo{person}{Jihyuk Park}, \bibinfo{person}{Yeji Han},
  {and} \bibinfo{person}{Uran Oh}.} \bibinfo{year}{2020}\natexlab{}.
\newblock \showarticletitle{Understanding Smartphone-based Online Shopping
  Experiences and Behaviors of Blind Users}.
\newblock \bibinfo{journal}{\emph{International journal of advanced smart
  convergence}} \bibinfo{volume}{9}, \bibinfo{number}{3}
  (\bibinfo{year}{2020}), \bibinfo{pages}{260--271}.
\newblock


\bibitem[Rapp(2021)]%
        {rapp2021search}
\bibfield{author}{\bibinfo{person}{Amon Rapp}.}
  \bibinfo{year}{2021}\natexlab{}.
\newblock \showarticletitle{In search for design elements: a new perspective
  for employing ethnography in human-computer interaction design research}.
\newblock \bibinfo{journal}{\emph{International Journal of Human--Computer
  Interaction}} \bibinfo{volume}{37}, \bibinfo{number}{8}
  (\bibinfo{year}{2021}), \bibinfo{pages}{783--802}.
\newblock


\bibitem[Sohaib and Kang(2017)]%
        {sohaib2017commerce}
\bibfield{author}{\bibinfo{person}{Osama Sohaib} {and} \bibinfo{person}{Kyeong
  Kang}.} \bibinfo{year}{2017}\natexlab{}.
\newblock \showarticletitle{E-commerce web accessibility for people with
  disabilities}.
\newblock In \bibinfo{booktitle}{\emph{Complexity in Information Systems
  Development}}. \bibinfo{publisher}{Springer}, \bibinfo{pages}{87--100}.
\newblock


\bibitem[Stangl et~al\mbox{.}(2020)]%
        {stangl2020person}
\bibfield{author}{\bibinfo{person}{Abigale Stangl},
  \bibinfo{person}{Meredith~Ringel Morris}, {and} \bibinfo{person}{Danna
  Gurari}.} \bibinfo{year}{2020}\natexlab{}.
\newblock \showarticletitle{" Person, Shoes, Tree. Is the Person Naked?" What
  People with Vision Impairments Want in Image Descriptions}. In
  \bibinfo{booktitle}{\emph{Proceedings of the 2020 chi conference on human
  factors in computing systems}}. \bibinfo{pages}{1--13}.
\newblock


\bibitem[Stangl et~al\mbox{.}(2018)]%
        {stangl2018browsewithme}
\bibfield{author}{\bibinfo{person}{Abigale~J. Stangl}, \bibinfo{person}{Esha
  Kothari}, \bibinfo{person}{Suyog~D. Jain}, \bibinfo{person}{Tom Yeh},
  \bibinfo{person}{Kristen Grauman}, {and} \bibinfo{person}{Danna Gurari}.}
  \bibinfo{year}{2018}\natexlab{}.
\newblock \showarticletitle{BrowseWithMe: An Online Clothes Shopping Assistant
  for People with Visual Impairments} \emph{(\bibinfo{series}{ASSETS '18})}.
  \bibinfo{publisher}{Association for Computing Machinery},
  \bibinfo{address}{New York, NY, USA}, \bibinfo{pages}{107–118}.
\newblock
\showISBNx{9781450356503}
\urldef\tempurl%
\url{https://doi.org/10.1145/3234695.3236337}
\showDOI{\tempurl}


\bibitem[Tateno et~al\mbox{.}(2020)]%
        {tateno2020method}
\bibfield{author}{\bibinfo{person}{Kiri Tateno}, \bibinfo{person}{Noboru
  Takagi}, \bibinfo{person}{Kei Sawai}, \bibinfo{person}{Hiroyuki Masuta},
  {and} \bibinfo{person}{Tatsuo Motoyoshi}.} \bibinfo{year}{2020}\natexlab{}.
\newblock \showarticletitle{Method for Generating Captions for Clothing Images
  to Support Visually Impaired People}. In \bibinfo{booktitle}{\emph{2020 Joint
  11th International Conference on Soft Computing and Intelligent Systems and
  21st International Symposium on Advanced Intelligent Systems (SCIS-ISIS)}}.
  IEEE, \bibinfo{pages}{1--5}.
\newblock


\bibitem[Wang et~al\mbox{.}(2021)]%
        {wang2021revamp}
\bibfield{author}{\bibinfo{person}{Ruolin Wang}, \bibinfo{person}{Zixuan Chen},
  \bibinfo{person}{Mingrui~Ray Zhang}, \bibinfo{person}{Zhaoheng Li},
  \bibinfo{person}{Zhixiu Liu}, \bibinfo{person}{Zihan Dang},
  \bibinfo{person}{Chun Yu}, {and} \bibinfo{person}{Xiang'Anthony' Chen}.}
  \bibinfo{year}{2021}\natexlab{}.
\newblock \showarticletitle{Revamp: Enhancing Accessible Information Seeking
  Experience of Online Shopping for Blind or Low Vision Users}. In
  \bibinfo{booktitle}{\emph{Proceedings of the 2021 CHI Conference on Human
  Factors in Computing Systems}}. \bibinfo{pages}{1--14}.
\newblock


\bibitem[Whitney(2020)]%
        {whitney2020teaching}
\bibfield{author}{\bibinfo{person}{Michael Whitney}.}
  \bibinfo{year}{2020}\natexlab{}.
\newblock \showarticletitle{Teaching Accessible Design: Integrating
  Accessibility Principles and Practices into an Introductory Web Design
  Course.}
\newblock \bibinfo{journal}{\emph{Information Systems Education Journal}}
  \bibinfo{volume}{18}, \bibinfo{number}{1} (\bibinfo{year}{2020}),
  \bibinfo{pages}{4--13}.
\newblock


\bibitem[Williams(2021)]%
        {j2021online}
\bibfield{author}{\bibinfo{person}{J'den~B. Williams}.}
  \bibinfo{year}{2021}\natexlab{}.
\newblock \bibinfo{booktitle}{\emph{The Online Apparel Shopping Experience of
  Blind Consumers}}.
\newblock \bibinfo{publisher}{North Carolina State University}.
\newblock


\bibitem[Yang et~al\mbox{.}(2015)]%
        {yang2015design}
\bibfield{author}{\bibinfo{person}{Huiqiao Yang}, \bibinfo{person}{Qijia Peng},
  \bibinfo{person}{Qin Gao}, {and} \bibinfo{person}{Pei-Luen Patrick~Rau}.}
  \bibinfo{year}{2015}\natexlab{}.
\newblock \showarticletitle{Design of a Clothing Shopping Guide Website for
  Visually Impaired People}. In \bibinfo{booktitle}{\emph{Cross-Cultural Design
  Methods, Practice and Impact}},
  \bibfield{editor}{\bibinfo{person}{P.L.Patrick Rau}} (Ed.).
  \bibinfo{publisher}{Springer International Publishing},
  \bibinfo{address}{Cham}, \bibinfo{pages}{253--261}.
\newblock
\showISBNx{978-3-319-20907-4}


\bibitem[Yu et~al\mbox{.}(2015)]%
        {yu2015retail}
\bibfield{author}{\bibinfo{person}{Hong Yu}, \bibinfo{person}{Sandra
  Tullio-Pow}, {and} \bibinfo{person}{Ammar Akhtar}.}
  \bibinfo{year}{2015}\natexlab{}.
\newblock \showarticletitle{Retail design and the visually impaired: A needs
  assessment}.
\newblock \bibinfo{journal}{\emph{Journal of Retailing and Consumer Services}}
  \bibinfo{volume}{24} (\bibinfo{year}{2015}), \bibinfo{pages}{121--129}.
\newblock


\bibitem[Yu et~al\mbox{.}(2018)]%
        {yu2018aesthetic}
\bibfield{author}{\bibinfo{person}{Wenhui Yu}, \bibinfo{person}{Huidi Zhang},
  \bibinfo{person}{Xiangnan He}, \bibinfo{person}{Xu Chen}, \bibinfo{person}{Li
  Xiong}, {and} \bibinfo{person}{Zheng Qin}.} \bibinfo{year}{2018}\natexlab{}.
\newblock \showarticletitle{Aesthetic-based clothing recommendation}. In
  \bibinfo{booktitle}{\emph{Proceedings of the 2018 world wide web
  conference}}. \bibinfo{pages}{649--658}.
\newblock


\bibitem[Zimmerman et~al\mbox{.}(2022)]%
        {zimmerman2022recentering}
\bibfield{author}{\bibinfo{person}{John Zimmerman}, \bibinfo{person}{Aaron
  Steinfeld}, \bibinfo{person}{Anthony Tomasic}, {and} \bibinfo{person}{Oscar
  J.~Romero}.} \bibinfo{year}{2022}\natexlab{}.
\newblock \showarticletitle{Recentering Reframing as an RtD Contribution: The
  Case of Pivoting from Accessible Web Tables to a Conversational Internet}. In
  \bibinfo{booktitle}{\emph{Proceedings of the 2022 CHI Conference on Human
  Factors in Computing Systems}}. \bibinfo{pages}{1--14}.
\newblock


\end{thebibliography}

\end{document}